\documentclass[a4paper,12pt]{article}

\usepackage{amssymb,amstext,amsmath}
\usepackage{graphicx}
\usepackage{color}
\usepackage{bm}
\usepackage{url}
\usepackage{tikz-cd}

\usepackage{authblk}

\usepackage{comment}

\usepackage[colorlinks=true,linkcolor=blue,citecolor=red]{hyperref}%

\def\q{{\bm{q}}}
\def\s{{\bm{s}}}
\def\x{{\bm{x}}}
\def\y{{\bm{y}}}
\def\pa{{\partial\Omega}}
\def\ve{\varepsilon}

\def\E{{\mathbb E}}
\def\P{{\mathbb P}}
\def\R{{\mathbb R}}
\def\Z{{\mathbb Z}}

\def\M{{\mathcal M}}

\def\erfc{\textrm{erfc}}
\def\erfcx{\textrm{erfcx}}

\title{Imperfect Diffusion-Controlled Reactions}
\author[1]{Denis S. Grebenkov}
\affil[1]{Laboratoire de Physique de la Mati\`ere Condens\'ee \\
CNRS -- Ecole Polytechnique, 91128 Palaiseau France\\ 
denis.grebenkov@polytechnique.edu}

\date{\today}

\begin{document}

\maketitle

\begin{abstract}
This chapter aims at emphasizing the crucial role of partial
reactivity of a catalytic surface or a target molecule in
diffusion-controlled reactions.  We discuss various microscopic
mechanisms that lead to imperfect reactions, the Robin boundary
condition accounting for eventual failed reaction events, and the
construction of the underlying stochastic process, the so-called
partially reflected Brownian motion.  We show that the random path to
the reaction event can naturally be separated into the transport step
toward the target, and the exploration step near the target surface
until reaction.  While most studies are focused exclusively on the
transport step (describing perfect reactions), the exploration step,
consisting is an intricate combination of diffusion-mediated jumps
between boundary points, and its consequences for chemical reactions
remain poorly understood.  We discuss the related mathematical
difficulties and recent achievements.  In particular, we derive a
general representation of the propagator, show its relation to the
Dirichlet-to-Neumann operator, and illustrate its properties in the
case of a flat surface.
\end{abstract}

\section*{Disclaimer}

Apart from the general introduction and motivation, the chapter
summarizes former results obtained by the author and his co-workers.
For this reason, the chapter presents the author's personal view on
the topic, whereas the bibliography is biased.

\section{Introduction}
\label{sec:intro}

Classical reaction kinetics describes the time evolution of spatially
homogeneous concentrations of reacting species via a set of coupled
nonlinear ordinary differential equations.  This description relies on
the assumption that the produced species move fast enough to maintain
homogeneous concentrations, and the limiting factor is the chemical
kinetics itself.  Marian von Smoluchowski was the first who emphasized
on the importance of the transport step by computing the macroscopic
reaction rate of small molecules $A$ diffusing toward a spherical
target $B$ of radius $R$ \cite{Smoluchowski17}.  If the reaction
occurs immediately upon the first encounter between the two molecules,
the rate is equal to the diffusive flux which is obtained the solving
the underlying diffusion equation on the concentration $c(\x,t)$ of
molecules $A$:
\begin{subequations}  \label{eq:diff}
\begin{eqnarray}  \label{eq:diff_eq}
\frac{\partial c(\x,t)}{\partial t} = D \Delta c(\x,t) &&  \qquad \textrm{diffusion in the bulk}~\Omega, \\
\label{eq:diff_ini}
c(\x,t=0) = c_0  && \qquad \textrm{uniform initial condition} , \\
\label{eq:diff_Dir}
c(\x,t)|_{\pa} = 0 && \qquad \textrm{vanishing on the target surface}~\pa , \\
\label{eq:diff_inf}
c(\x,t)|_{|\x| \to \infty} = c_0  && \qquad \textrm{regularity condition at infinity}, 
\end{eqnarray}
\end{subequations}
where $D$ is the diffusion coefficient of molecules $A$ (in units
m$^2$/s), $c_0$ is their initial concentration (in units mol/m$^3$),
the confining domain $\Omega = \{ \x \in \R^3 ~:~ |\x| > R\}$ is the
bulk space outside the target, $\pa = \{ \x \in \R^3 ~:~ |\x| = R\}$
is the reactive surface of the target, and $\Delta =
\partial^2/\partial x^2 + \partial^2/\partial y^2 +
\partial^2/\partial z^2$ is the Laplace operator governing diffusion.
Here and throughout the text, we assume that the target $B$ is
immobile while the molecules $A$ can be treated as independent and
point-like.  Although Smoluchowski considered both molecules to be
mobile and of finite size (in which case $D$ and $R$ are respectively
the sum of diffusivities and of radii of two molecules), the
equivalence between two settings is not valid in general.

The diffusion equation (\ref{eq:diff_eq}) states that the time
evolution of the concentration (the left-hand side) is caused
exclusively by diffusive motion (the right-hand side), whereas the
Dirichlet boundary condition (\ref{eq:diff_Dir}) ensures that any
molecule $A$ arriving at the boundary of the target $B$ immediately
reacts, i.e., it is transformed into another molecule $A'$,
\begin{equation}  \label{eq:reaction}
A + B \longrightarrow A' + B ,
\end{equation}
which diffuses away.  This is the basic catalytic reaction, in which
the target molecule $B$ is a catalyst, needed to initiate the chemical
transformation of $A$ into $A'$.  Since the molecule $A$, being
transformed into $A'$ upon the contact, disappears at the target
surface, the concentration $c(\x,t)$ vanishes on $\pa$.  Finally, the
regularity condition (\ref{eq:diff_inf}) claims that the concentration
(infinitely) far away from the target remains unaffected and equal to
$c_0$.

Smoluchowski provided the exact solution of the partial differential
equation (\ref{eq:diff}) and deduced the diffusive flux of molecules
$A$ onto the target:
\begin{equation}  \label{eq:J_Smol}
J_{\rm S}(t) = \int\limits_{|\x| = R} d\s \, \left. \biggl(- D\frac{\partial c(\x,t)}{\partial n}\biggr) \right|_{\pa} =
4\pi c_0 D R \biggl(1 + \frac{R}{\sqrt{\pi D t}} \biggr),
\end{equation}
where $\partial/\partial n$ is the normal derivative at the boundary,
oriented outward the domain (i.e., toward to center of the target in
this case).  In the long-time limit, the second term vanishes, and one
gets the steady-state reaction rate, $J_{\rm S}(\infty) = 4\pi c_0
DR$, which is proportional to the diffusion coefficient $D$ and to the
radius $R$ of the target.  The seminal work by Smoluchowski focused
exclusively onto the transport step as a limiting factor, considering
{\it perfect} immediate reactions upon the first encounter.  This
setting was later called ``diffusion-limited reactions'', in contrast
to conventional ``kinetics-limited reactions'' \cite{Rice85}.  In
numerous following studies, the basic diffusion problem
(\ref{eq:diff}) was extended in various directions, in particular, by
replacing the exterior of a spherical target by an arbitrary Euclidean
domain $\Omega \subset \R^d$ \cite{Carslaw,Crank}, by considering one
or multiple targets on the otherwise inert impenetrable boundary
\cite{Redner,Holcman14,Holcman,Grebenkov16c}, by replacing the Laplace
operator (ordinary diffusion) by a general second-order elliptic
differential operator \cite{Bass} or a general Fokker-Planck operator
\cite{Risken}, by introducing bulk reactivity
\cite{Yuste13,Meerson15,Grebenkov17d}, trapping events
\cite{Bouchaud90,Metzler00}, or intermittence
\cite{Benichou10,Benichou11,Rupprecht12a,Rupprecht12b}.
However, the focus on the transport step till the first encounter with
the target, expressed via Eq. (\ref{eq:diff_Dir}), still remains
the dominant paradigm nowadays.

In practice, chemical reactions always combine the transport step
until an encounter and the reaction step.  While in some situations
one of these steps can be much longer than the other, it is important
to consider them together, as the impact of a seemingly ``irrelevant''
step may still be crucial, as we discuss below.  Collins and Kimball
proposed to describe {\it imperfect} reactions on the target surface
$\pa$ by replacing the Dirichlet boundary condition
(\ref{eq:diff_Dir}) by the Robin boundary condition (also known as
Fourier, radiation or third boundary condition):
\begin{equation}  \label{eq:diff_Rob}
\biggl(-D \frac{\partial c(\x,t)}{\partial n}\biggr)\biggl|_{\pa} \biggr. = \kappa \, c(\x,t)|_{\pa} ,
\end{equation}
where $\kappa$ (in units m/s) is called the {\it reactivity}
\cite{Collins49}.  Although $\kappa$ can in general be any nonnegative
function of a boundary point, we focus on the practically relevant
case of a constant $\kappa$ (we also discuss below the mixed
Robin-Neumann boundary condition when $\kappa$ is a piecewise constant
function).

The relation (\ref{eq:diff_Rob}) states that the diffusive flux
density of molecules at the target surface (the left-hand side) is
proportional to the concentration of the molecules at the target (the
right-hand side), $\kappa$ being the proportionality coefficient.  It
is important to stress that the diffusive flux density is a
macroscopic quantity that describes the average difference between the
diffusive flux toward the target and the diffusive flux from the
target back to the bulk.  In the limit $\kappa = 0$, the diffusive
flux density vanishes at the boundary $\pa$, meaning that, on average,
the number of molecules $A$ arriving from the bulk onto the target
surface is equal to the number of molecules diffusing from the target
back to the bulk.  As a consequence, none of these molecules actually
react on the target.  So the limit $\kappa =0$ and the corresponding
Neumann boundary condition describes a chemically inactive (inert)
boundary that just confines the molecules in a prescribed spatial
region, with no reaction.  When $\kappa > 0$, there is a net
difference between the diffusive flux toward the target and that from
the target, and this difference is precisely the flux of reacted
molecules.  In the limit $\kappa\to\infty$, one retrieves the
Dirichlet boundary condition (\ref{eq:diff_Dir}) by dividing
Eq. (\ref{eq:diff_Rob}) by $\kappa$ and taking the limit.  Varying
$\kappa$ from zero to infinity allows one to change the chemical
reactivity of the target from inert to perfectly reactive.

Collins and Kimball solved the diffusion problem (\ref{eq:diff}) with
the Robin boundary condition (\ref{eq:diff_Rob}) and found the
macroscopic reaction rate \cite{Collins49}
\begin{equation}  \label{eq:J_Collins}
J_{\rm CK}(t) = \frac{4\pi c_0 D R}{1 + \frac{D}{\kappa R}} \biggl(1 + \frac{\kappa R}{D} \, \erfcx\bigl( (1/R + \kappa/D)\sqrt{Dt}\bigr) \biggr),
\end{equation}
where $\erfcx(z)$ is the scaled complementary error function (see
Ref. \cite{Traytak07} for an extension to distinct reactivities on two
complementary caps of a sphere and Ref. \cite{Qian06} for a
perturbative approach for a dilute suspension of multiple partially
reactive spheres).  In the diffusion-limited case $\kappa\to \infty$,
one recovers the Smoluchowski rate $J_{\rm S}(t)$, whereas the
reaction-limited case $\kappa\to 0$ gives the classical limit $J_{\rm
reac} = 4\pi R^2 \, c_0 \kappa$, i.e., the rate is proportional to the
surface area of the target and to the reactivity $\kappa$ but is
independent of the diffusion coefficient $D$.  One can combine these
limiting expressions to rewrite the steady-state limit of
Eq. (\ref{eq:J_Collins}) as
\begin{equation}  \label{eq:J_Collins2}
\frac{1}{J_{\rm CK}(\infty)} = \frac{1}{J_{\rm S}(\infty)} + \frac{1}{J_{\rm reac}} \,.
\end{equation}
Understanding the inverse of the flux as a ``resistance'', one gets a
simple electric interpretation of the Collins-Kimball rate $J_{\rm
CK}(\infty)$ as the serial connection of two ``resistances''
characterizing the ``difficulty'' to access the target (the transport
step) and the ``difficulty'' to react on the target (the reaction
step) \cite{Sapoval94}.  Alternatively, one can think of this relation
as the sum of two characteristic times of the underlying steps.  Note
that a similar additivity law was established for the mean reaction
time \cite{Grebenkov17a,Grebenkov17b}.  Chemical reactions for which
both the transport and reaction steps are relevant are sometimes
called {\it diffusion-influenced}, {\it diffusion-mediated}, or {\it
diffusion-controlled}.

The macroscopic boundary condition (\ref{eq:diff_Rob}) can describe
various {\it microscopic} mechanisms of imperfect reactions.

(i) A molecule needs to overcome an energetic activation barrier for
reaction to occur \cite{Hanggi90}; in this setting, the reaction
(\ref{eq:reaction}) can be written more accurately as
\begin{equation}  \label{eq:reaction2}
\begin{tikzcd}[every arrow/.append style={shift left}]
A + B \arrow{r}{k_{\rm on}} & AB \arrow{l}{{k_{\rm off}}} \overset{k_{\rm reac}}{\longrightarrow} A' + B ,
\end{tikzcd}
\end{equation}
where $AB$ is a metastable complex which can either result in the
production of $A'$ (if the activation barrier is overpassed, with the
rate $k_{\rm reac}$), or be dissociated back to $A$ and $B$ (with the
dissociation rate $k_{\rm off}$).  The reactivity is thus related to
the height of the activation barrier.  For instance, the reactivity
$\kappa$ of a spherical target can be directly related to the
association rate $k_{\rm on}$ (in units m$^3$/s/mol) as $k_{\rm on} =
4\pi R^2 \, \kappa \, N_{\rm av}$, where $N_{\rm av} \simeq 6\cdot
10^{23}$~1/mol is the Avogadro number, and $4\pi R^2$ is the surface
area of the target \cite{Shoup82,Lauffenburger}.  This description was
shown to be important for partial diffusion-controlled recombination
\cite{Sano79,Sano81}.  The reversible binding in (\ref{eq:reaction2})
is also the key element for coupling bulk diffusion to other
diffusion-reaction processes on the target.  For instance, such a
coupling was employed to describe the flux of receptors across the
boundary between the dendrite and its spines \cite{Bressloff08}.

(ii) A molecule needs to overcome an entropic barrier
\cite{Zhou91,Reguera06} if the reaction is understood as an escape
from a confining domain through a small opening region on the boundary
(e.g., ion channels or aquaporins on the membrane of a living cell).
In this case, the reactivity can be related to the geometric shape of
the opening region (see Ref. \cite{Grebenkov17a} for further
discussion).

(iii) There is a stochastic gating when the target can randomly switch
between ``open'' and ``closed'' states, implying that the diffusing
molecule can either go through the open gate, or be stopped at the
closed gate and thus reflected back to resume its diffusion
\cite{Benichou00,Reingruber09,Bressloff17}; the same stochastic
mechanism is relevant for enzymatic reactions when an enzyme can
randomly change its conformational state to be active or passive.  In
both cases, the reactivity is related to the switching rates or to the
probability of finding the gate open or the enzyme active.

(iv) When the target is an inert surface covered by small reactive
patches, the diffusing molecule can either hit a patch and react
immediately, or be reflected on an inert part and resume its diffusion
until the next arrival on the target.  Homogenizing such microscopic
reactivity heterogeneities by setting an effective finite reactivity
$\kappa$ uniformly on the target is often used to facilitate the
analysis (given that the diffusion equation with multiple targets is a
much more complicated problem, for both analytical and numerical
computations).  For instance, for a spherical target covered uniformly
by $N$ disks of radius $a$, Berg and Purcell obtained \cite{Berg77}
$\kappa = D Na/(\pi R^2)$ (see further extensions in
Refs. \cite{Shoup81,Zwanzig91}).

(v) For larger scale problems (e.g., animal foraging), the finite
reactivity can model a ``recognition'' step when a particle or a
species may need to recognize the target \cite{Grebenkov10a}.

The above (incomplete) list of microscopic mechanisms urges for
considering imperfect reactions.  We also mention that the Robin
boundary condition (\ref{eq:diff_Rob}) can describe permeation across
a membrane (in which case $\kappa$ is called permeability)
\cite{Powles92,Sapoval02,Grebenkov05}, impedance of an electrode
\cite{Sapoval94,Filoche99,Grebenkov06}, surface relaxation in nuclear
magnetic resonance
\cite{Brownstein79,Grebenkov07,Grebenkov16d}, etc.

According to Eq. (\ref{eq:J_Collins2}), the macroscopic reaction rate
results from a balance between the transport and the reaction steps,
whereas the diffusion coefficient $D$ and the reactivity $\kappa$ play
the roles of respective weights in the Robin boundary condition
(\ref{eq:diff_Rob}).  The ratio of these quantities naturally defines
the {\it reaction length} \cite{Sapoval94}, 
\begin{equation}
\Lambda = D/\kappa, 
\end{equation}
which has to be compared to {\it geometric length scales} of the
problem (e.g., the radius of the target $R$ in our spherical example).
In particular, the diffusion-limited and reaction-limited cases
correspond to conditions $\Lambda \ll R$ and $\Lambda \gg R$,
respectively.

When molecular diffusion occurs in a structurally complex environment
(e.g., an overcrowded cytoplasm or a multiscale porous medium), both
the transport and reaction steps strongly depend on the geometrical
structure of the environment.  While the transport step is relatively
well understood (see Refs. \cite{Condamin07,Benichou10b,Benichou14}
and references therein), the more sophisticated reaction step was most
often just ignored.  In the next section, we summarize the main steps
of a mathematical construction of the stochastic process allowing one
to investigate the reaction step.

\section{Partially reflected Brownian motion}
\label{sec:proba}

The macroscopic formulation of diffusion-reaction processes in terms
of partial differential equations such as Eq. (\ref{eq:diff}) admits
an equivalent microscopic probabilistic interpretation in terms of
random trajectories of appropriate stochastic processes
\cite{Freidlin,Schuss80,Bass,Stroock71}.  From the mathematical point
of view, the Dirichlet boundary condition (\ref{eq:diff_Dir}) is the
easiest to deal with.  In fact, for a given continuous stochastic
process $X_t$, it is sufficient to define the first passage time
$\tau_0$ to the target surface $\pa$,
\begin{equation}  \label{eq:tau0}
\tau_0 = \inf\{ t > 0 ~:~ X_t \in \pa\}, 
\end{equation}
and then to stop the process at this time.  For instance, the solution
of the steady-state diffusion equation, $\Delta u(\x) = 0$, subject to
the Dirichlet boundary condition, $u(\x)|_{\pa} = \psi(\x)$, is $u(\x)
= \E\{ \psi(X_{\tau_0}) | X_0 = x\}$, where $\E$ denotes the
(conditional) expectation and $\psi(\x)$ is a prescribed function on
the boundary.  In other words, one can think of the solution $u(\x)$
as the average of the function $\psi(\x)$ evaluated at random boundary
points $X_{\tau_0}$ at which the stochastic process $X_t$ (started
from $x$) has arrived onto the boundary $\pa$ for the first time.  In
this way, the presence of a perfectly absorbing boundary can be
accounted for via the first passage time $\tau_0$, {\it with no
modification to the stochastic process itself}.

In contrast, accounting for an inert or partially reactive boundary
requires to modify the stochastic process.  In fact, when a molecule
arrives onto an inert boundary, its motion across this boundary should
be prohibited.  Physically, one can think of applying a very strong
potential, localized in an infinitesimal vicinity of the boundary, to
force the molecule moving back into the bulk.  In mathematical terms,
the so-called {\it reflected Brownian motion} can be defined as the
solution of the stochastic Skorokhod equation \cite{Freidlin}.  This
is a standard probabilistic way of implementing the Neumann boundary
condition for an inert boundary.  Once the reflected Brownian motion
is constructed, one can implement eventual reactions to treat the
Robin boundary condition
\cite{Ma90,Papanicolaou90,Milshtein95,Bass08,Schuss}.  While the related
mathematical theory is well developed, its details are beyond the
scope of this chapter.  

\begin{figure}
\begin{center}
\includegraphics[width=110mm]{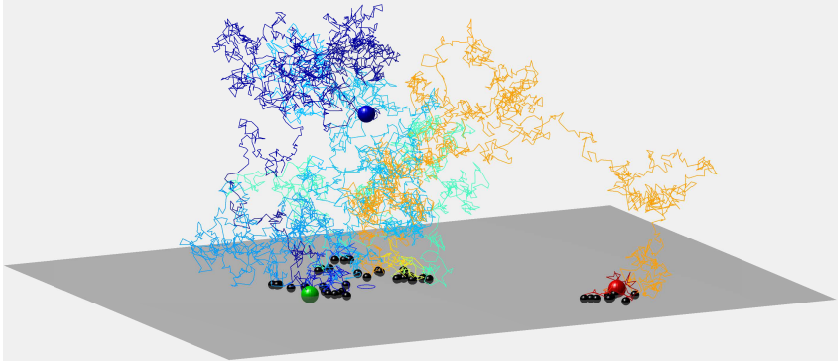}
\end{center}
\caption{
A simulated trajectory of partially reflected Brownian motion with
$\Lambda = 10$ and $a = 0.05$.  Blue, green and red balls show
respectively the starting point $(0,0,2)$, the first arrival point,
and the reaction point, whereas small black balls indicate reflection
events.  Diffusion-mediated jumps are drawn by different colors.
Length units are arbitrary.  We chose this continuous-space simulation
instead of a lattice random walk for a better visualization.}
\label{fig:traj}
\end{figure}

The easier and physically more appealing approach consists in modeling
the diffusion process as a discrete-time random walk on a lattice
$\Z^d$ with the inter-site distance $a$.  In a bulk site, the molecule
jumps with the probability $1/(2d)$ to one of the neighboring sites.
When the molecule sits at a boundary site, it can either react with
the probability
\begin{equation}  \label{eq:qa}
q_a = (1 + \Lambda/a)^{-1}, 
\end{equation}
or move back to a neighboring bulk site with the probability $1-q_a$
and continue the random walk until the next arrival onto the boundary
\cite{Filoche99,Grebenkov03}.  The trajectory of a molecule near a
partially reactive boundary can thus be seen as a sequence of
independent diffusion-mediated jumps between boundary points, each
jump being a random walk in the bulk that starts from the closest bulk
neighbor of a boundary site and terminates at another boundary site
upon the first arrival onto the boundary (Fig. \ref{fig:traj}).  This
process is called {\it partially reflected random walk}, whereas the
statistical properties of the jumps are described by the so-called
Brownian self-transport operator \cite{Filoche99,Grebenkov03}.  One
naturally recovers the two limiting cases: certain reaction ($q_a =
1$) for a perfectly reactive surface ($\Lambda = 0$ or $\kappa =
\infty$) and certain reflection ($q_a = 0$) for an inert surface
($\Lambda = \infty$ or $\kappa = 0$).

Since the reaction event and jumps are independent, the random number
$\eta$ of jumps until reaction follows a geometric distribution: $\P\{
\eta = n \} = q_a (1-q_a)^n$ ($n = 0,1,2,\ldots$).  In particular, the
mean number of jumps, $\E\{ \eta \} = (1-q_a)/q_a = \Lambda/a$, is
proportional to the reaction length $\Lambda$ and inversely
proportional to the lattice mesh size $a$.  When $a$ is small, it is
thus convenient to consider a rescaled variable $\chi =
\eta a$ which follows an exponential distribution:
\begin{equation}  \label{eq:chi}
\P\{\chi \geq x\} = \P\{ \eta \geq x/a \} = (1 - q_a)^{x/a} = (1 + a/\Lambda)^{-x/a} \approx e^{-x/\Lambda}.
\end{equation}
From the probabilistic point of view, consecutive individual trials
after each jump with probability $q_a$ are fully equivalent to saying
that the trajectory is stopped when the number of realized jumps,
multiplied by $a$, exceeds an independent exponentially distributed
random variable $\chi$.

Each jump started at the distance $a$ from the boundary has high
chances to return to the boundary within the distance $a$ to the
starting point.  In other words, each jump explores the boundary at a
typical distance of the order of $a$ (we emphasize that the {\it mean}
exploration distance can be infinite, see Sec. \ref{sec:half}).  As a
consequence, the whole trajectory from the first arrival onto the
boundary until reaction, being a sequence of $\Lambda/a$ jumps on
average, explores the surface up to a typical distance $\Lambda$.
These qualitative arguments provide the geometric meaning of the
reaction length $\Lambda$ as a typical size of the surface region
around the first arrival point explored until reaction (see
Sec. \ref{sec:half} for precise statements).

The above lattice-based construction of partially reflected random
walks employs the mesh distance $a$.  In the limit $a\to 0$, each jump
starts closer to the boundary and thus explores shorter distances, but
the number of jumps increases (given that $q_a \to 0$) so that these
effects compensate each other.  One can thus expect that such
partially reflected trajectories on a lattice converge to a
well-defined continuous limit as $a \to 0$ (in the same way as
ordinary random walks converge to Brownian motion).  This limit that
we call {\it partially reflected Brownian motion} (PRBM)
\cite{Grebenkov06a}, is defined as reflected Brownian motion stopped
at the random time
\begin{equation}  \label{eq:tau_Lambda}
\tau_\Lambda = \inf\{ t>0 ~:~ \ell_t > \chi\}, 
\end{equation}
where $\ell_t$ is the local time process of the reflected Brownian
motion on the boundary, which is a continuous analog of the rescaled
number of jumps on the boundary ($\eta a$) up to time $t$, and an
independent random variable $\chi$ was defined by Eq. (\ref{eq:chi}).
The reaction length $\Lambda$ appears in the exponential law
(\ref{eq:chi}) and thus parameterizes the family of PRBMs.  In the
limit $\Lambda = 0$, the exponential distribution for $\chi$
degenerates to the Dirac distribution at $0$, so that
Eq. (\ref{eq:tau_Lambda}) is reduced to $\tau_0 = \inf\{ t>0 ~:~
\ell_t > 0\}$, i.e., $\tau_0$ is the first moment when the local time
process becomes strictly positive that occurs on the first arrival
onto the boundary, and one retrieves Eq. (\ref{eq:tau0}).

In spite of apparent similarities between discrete-space and
continuous-space partially reflected diffusions, there is a peculiar
difference in their construction.  In the discrete-space approach, one
considers a sequence of diffusion-mediated jumps, in which each jump
terminates upon the first arrival on the boundary.  In other words,
the boundary is treated as perfectly absorbing for each jump, whereas
the partial reactivity is controlled by the number of jumps.  In
contrast, in the continuous-space approach, the underlying process is
the reflected Brownian motion so that the boundary is treated as fully
reflecting, whereas the partial reactivity is again controlled by the
(rescaled) number of jump (i.e., the local time process).

The introduction of a small distance $a$ to restart each jump was
crucial to avoid an immediate return to the boundary.  In contrast,
discrete space and time random walks on a lattice can be easily
replaced by continuous trajectories of Brownian motion
\cite{Grebenkov06a,Grebenkov07a}.  While we focused on partially
reflected Brownian motion governed by the Laplace operator, one can
construct much more general partially reflected diffusions governed by
elliptic second-order differential operators \cite{Schuss}.  In
particular, one can consider Markovian jump processes generated by the
Euler discretization scheme of a more general stochastic process in
the bulk \cite{Singer08}.  Once the generated trajectory crosses the
boundary, it can be either terminated with probability $q_a$, or
reflected back.  In our notations, the probability $q_a$ obtained by
Singer {\it et al.} reads
\begin{equation}  \label{eq:qa_Singer}
q_a = \sqrt{\pi} \kappa \sqrt{\Delta t}/\sqrt{2D} = (\sqrt{\pi}/2) \kappa a/D = (\sqrt{\pi}/2) a/\Lambda, 
\end{equation}
with $a = \sqrt{2D\Delta t}$, where $\Delta t$ is the time step.  This
expression is close to the formula (\ref{eq:qa}), which for small
$a/\Lambda$ yield $q_a \approx a/\Lambda$.  The extra factor
$\sqrt{\pi}/2 \approx 0.89$ is related to the use of Gaussian jumps in
the Euler scheme instead of discrete-space random walks.  In contrast
to Eq. (\ref{eq:qa_Singer}), the formula (\ref{eq:qa}) is applicable
for any reactivity, i.e. $a/\Lambda$ does not need to be a small
parameter.

In the next section, we discuss the propagator of PRBM and show the
impact of the reactivity onto various statistics relevant to
diffusion-controlled reactions.

\section{General representation of the propagator}

In this section, we derive a general representation of the propagator
$G_\Lambda(\x_0,\x;t)$ of partially reflected Brownian motion, i.e.,
the probability density of finding a molecule started at a point
$\x_0$ at time $0$ in a vicinity of a point $\x$ at time $t$ in the
presence of a partially reactive boundary $\pa$ characterized by the
reaction length $\Lambda$.  The propagator satisfies the following
equations
\begin{subequations}
\begin{eqnarray}
\frac{\partial G_\Lambda(\x_0,\x;t)}{\partial t} - D \Delta  G_\Lambda(\x_0,\x;t) &=& 0  \qquad (\x \in \Omega), \\
G_\Lambda(\x_0,\x;t=0) &=& \delta(\x - \x_0) , \\  \label{eq:G_Robin}
\biggl(\Lambda \frac{\partial}{\partial n} + 1\biggr)  G_\Lambda(\x_0,\x;t) &=& 0  \qquad (\x \in \pa),  
\end{eqnarray}
\end{subequations}
where $\delta(\x-\x_0)$ is the Dirac distribution (if the domain
$\Omega$ is unbounded, these equations should be completed by the
regularity condition at infinity: $G_\Lambda(\x_0,\x;t) \to 0$ as
$|\x|\to\infty$).  To avoid technical details, we assume that the
boundary $\pa$ is smooth.

The propagator can be decomposed into two parts: the contribution of
direct trajectories from $\x_0$ to $\x$ without touching the reactive
surface, $G_0(\x_0,\x;t)$, and the contribution of trajectories that
hit this surface at a point $\s_1\in\pa$ at an intermediate time $0 <
t_1 < t$:
\begin{equation}  \label{eq:Gprob1}
G_\Lambda(\x_0,\x;t) = G_0(\x_0,\x;t) + \int\limits_\pa d\s_1 \int\limits_0^t dt_1 \, j_0(\x_0,\s_1;t_1) \, G_\Lambda(\s_1,\x;t-t_1),
\end{equation}
where
\begin{equation}
j_0(\x_0,\s_1;t_1) = \biggl(- D \frac{\partial}{\partial n} G_0(\x_0,\x;t_1)\biggr)_{\x = \s_1} 
\end{equation}
is the diffusive flux density at time $t_1$ at a point $\s_1$ of the
{\it perfectly} reactive surface (i.e., the probability density of the
first arrival in a vicinity of $\s_1$ at time $t_1$).  This density
describes the transport step and is independent of $\Lambda$.  Next,
employing the reversal symmetry of the propagator,
$G_\Lambda(\x_0,\x;t) = G_\Lambda(\x,\x_0;t)$, which is also valid
when $\x_0$ is a boundary point for $\Lambda > 0$, one can represent
$G_\Lambda(\s_1,\x;t)$ in Eq. (\ref{eq:Gprob1}) using again
Eq. (\ref{eq:Gprob1}) to obtain
\begin{eqnarray}    \nonumber
G_\Lambda(\x_0,\x;t) &=& G_0(\x_0,\x;t) + \int\limits_\pa d\s_1 \int\limits_\pa d\s_2 \int\limits_0^t dt_1 \int\limits_{t_1}^t dt_2 
\, j_0(\x_0,\s_1;t_1) \\    \label{eq:Gprob}   
&& \times  G_\Lambda(\s_1,\s_2;t_2-t_1) \, j_0(\x,\s_2;t-t_2) .
\end{eqnarray}
This relation expresses the propagator $G_\Lambda(\x_0,\x;t)$ in the
whole domain in terms of the propagator $G_\Lambda(\s_1,\s_2;t)$ from
one boundary point to another via bulk diffusion.  The second term in
Eq. (\ref{eq:Gprob}) has a simple probabilistic interpretation: a
molecule reaches the boundary for the first time at $t_1$, performs
partially reflected Brownian motion (with eventual reaction) over time
$t_2-t_1$ and, if not reacted during this time, diffuses to the bulk
point $\x$ during time $t-t_2$ without hitting the reactive surface.
We emphasize that the last step would not be possible in the perfectly
reactive case because the propagator $G_0(\x_0,\x;t)$ is zero when
$\x$ is a boundary point.  We also stress that Eq. (\ref{eq:Gprob}) is
valid if $\x = \s\in \pa$ is a boundary point.  In this case,
$G_0(\x_0,\s;t) = 0$, while $j_0(\s,\s_2;t-t_2) = \delta(\s-\s_2)
\delta(t-t_2)$, so that the integrals over $\s_2$ and $t_2$ are
removed, reducing Eq. (\ref{eq:Gprob}) to
\begin{equation}  \label{eq:Gprob_s}
G_\Lambda(\x_0,\s;t) = \int\limits_\pa d\s_1 \int\limits_0^t dt_1 \, j_0(\x_0,\s_1;t_1) \, G_\Lambda(\s_1,\s;t-t_1).
\end{equation}
This relation justifies the qualitative separation of the
diffusion-reaction process into two steps: the transport step
(described by $j_0(\x_0,\s_1;t_1)$) and the reaction step (described
by $G_\Lambda(\s_1,\s;t-t_1)$ or related quantities, see below).  We
stress, however, that the reaction step involves intricate diffusion
process near the partially reactive surface.  In addition to the new
conceptual view onto partially reflected Brownian motion, the
representations (\ref{eq:Gprob}, \ref{eq:Gprob_s}) can be helpful for
a numerical computation of the propagator because only the
boundary-to-boundary transport $G_\Lambda(\s_1,\s_2;t)$ needs to be
evaluated.  Note also that the time convolution can be removed by
passing to the Laplace domain, in which Eq. (\ref{eq:Gprob}) becomes
\begin{eqnarray}   \label{eq:Gprob_Laplace}
\tilde{G}_\Lambda(\x_0,\x;p) &=& \tilde{G}_0(\x_0,\x;p) \\    \nonumber
&+& \int\limits_\pa d\s_1 \int\limits_\pa d\s_2  \, \tilde{j}_0(\x_0,\s_1;p)  \,
\tilde{G}_\Lambda(\s_1,\s_2;p) \, \tilde{j}_0(\x,\s_2;p) ,
\end{eqnarray}
where tilde denotes the Laplace transform: $\tilde{f}(p) =
\int\nolimits_0^\infty dt \, e^{-pt} \, f(t)$.

\subsection*{Relation to the Dirichlet-to-Neumann operator}

The Laplace-transformed propagator $\tilde{G}_\Lambda(\s_1,\s_2;p)$
turns out to be the resolvent of the {\it Dirichlet-to-Neumann
operator} $\M_p$ for the modified Helmholtz equation: to a given
function $f$ on the boundary $\pa$ of a confining domain $\Omega
\subset \R^d$, the operator $\M_p$ associates another function on the
boundary,
\begin{equation}
\M_p f = \biggl(\frac{\partial u(\x;p)}{\partial n}\biggr)_{|\x\in\pa}, 
\end{equation}
where $u(\x;p)$ is the solution of the Dirichlet boundary value
problem:
\begin{subequations}
\begin{eqnarray}
(p - D \Delta) u(\x;p) &=& 0 \hskip 12mm (\x\in\Omega), \\
u(\x;p)_{|\pa} &=& f(\x;p) \quad (\x\in\pa) .
\end{eqnarray}
\end{subequations}
For instance, if $f(\x;p)$ is understood as a source of molecules on
the boundary $\pa$ emitted into the reactive bulk, then the operator
$\M_p$ gives their flux density on that boundary.  Note that there is
a family of operators parameterized by $p$ (or $p/D$).  While the
Dirichlet-to-Neumann operator was conventionally studied for the
Laplace equation (i.e., $p = 0$), the above extension to the modified
Helmholtz equation is straightforward.

One can use the Dirichlet-to-Neumann operator to represent the
Laplace-transformed propagator $\tilde{G}_\Lambda(\x_0,\x;p)$ which
satisfies for each fixed $\x_0 \in \Omega$ the modified Helmholtz
equation:
\begin{subequations}
\begin{eqnarray}
(p - D \Delta) \tilde{G}_\Lambda(\x_0,\x;p) &=& \delta(\x - \x_0)  \quad (\x\in\Omega), \\
\biggl(\Lambda \frac{\partial}{\partial n} + 1 \biggr) \tilde{G}_\Lambda(\x_0,\x;p) &=& 0 \quad (\x \in \pa) .
\end{eqnarray}
\end{subequations}
To get rid off $\delta(\x-\x_0)$, one can search the solution in the
form 
\begin{equation}   \label{eq:g_reg}
\tilde{G}_\Lambda(\x_0,\x;p) = \tilde{G}_0(\x_0,\x;p) + \tilde{g}_\Lambda(\x_0,\x;p), 
\end{equation}
where the unknown regular part $\tilde{g}_\Lambda(\x_0,\x;p)$
satisfies
\begin{subequations}
\begin{eqnarray}
(p - D \Delta) \tilde{g}_\Lambda(\x_0,\x;p) &=& 0  \quad (\x\in\Omega), \\  \label{eq:auxil2}
\biggl(\Lambda \frac{\partial}{\partial n} + 1 \biggr) \tilde{g}_\Lambda(\x_0,\x;p) &=& 
\underbrace{- \Lambda \biggl(\frac{\partial}{\partial n} \tilde{G}_0(\x_0,\x;p)\biggr)}_{=(\Lambda/D) \tilde{j}_0(\x_0,\x;p)} \quad (\x \in \pa) .
\end{eqnarray}
\end{subequations}
 
Suppose that we have solved this problem and found that the solution
$\tilde{g}_\Lambda(\x_0,\x;p)$ on the boundary $\pa$ is equal to some
function $f(\x;p)$.  Once we know $f(\x;p)$, we can simply reconstruct
$\tilde{g}_\Lambda(\x_0,\x;p)$ as the solution of the corresponding
Dirichlet problem.  Applying then the Dirichlet-to-Neumann operator to
$f(\x;p)$, we can express the normal derivative of
$\tilde{g}_\Lambda(\x_0,\x;p)$.  Summarizing these steps, one can
express the restriction of the solution $\tilde{g}_\Lambda(\x_0,\x;p)$
onto the boundary as
\begin{equation}
\tilde{g}_\Lambda(\x_0,\s ; p) = \bigl(\Lambda \M_p + I\bigr)^{-1} \, \frac{\Lambda}{D} \tilde{j}_0(\x_0,\s;p) \qquad (\s\in\pa),
\end{equation}
where $I$ is the identity operator.  Finally, if the starting point
$\x_0 = \s_0$ lies on the boundary, one has $\tilde{j}_0(\s_0,\s;p) =
\delta(\s - \s_0)$, the Dirichlet propagator $\tilde{G}_0(\s_0,\x;p)$
vanishes in Eq. (\ref{eq:g_reg}), and we get
\begin{equation}
D \tilde{G}_\Lambda(\s_0,\s;p) = \bigl(\M_p + I/\Lambda\bigr)^{-1}  \delta(\s - \s_0) \qquad (\s_0,\s\in\pa).
\end{equation}
We conclude that $D \tilde{G}_\Lambda(\s_0,\s;p)$ is the resolvent of
the Dirichlet-to-Neumann operator $\M_p$.  In particular, one can
rewrite Eq. (\ref{eq:Gprob_Laplace}) in the form of a scalar product
between two functions on the boundary
\begin{equation}
\tilde{G}_\Lambda(\x_0,\x;p) = \tilde{G}_0(\x_0,\x;p) + \frac{\Lambda}{D} \biggl( \tilde{j}_0(\x_0,\cdot;p)  
\, \cdot \, (I + \Lambda \M_p)^{-1} \tilde{j}_0(\x,\cdot;p) \biggr)_{L_2(\pa)} .
\end{equation}
Remarkably, all the ``ingredients'' of this formula correspond to the
Dirichlet condition on a perfectly reactive boundary, and only the
reaction length $\Lambda$ keeps track of partial reactivity.  Note
that this is a significant extension of the formula for the total
steady-state flux derived in Ref. \cite{Grebenkov06}.  This profound
connection (that was earlier established for the conventional setting
of the Laplace equation \cite{Grebenkov06a,Grebenkov06}) allows one to
express many probabilistic quantities through the spectral properties
of the Dirichlet-to-Neumann operator.

\subsection*{Reaction time distribution and spread harmonic measure}

The propagator determines many quantities often considered in the
context of diffusion-controlled reactions: the survival probability up
to time $t$, the reaction time distribution, the distribution of the
reaction point (at which reaction occurs), etc.  For instance, the
diffusive flux density at a partially reactive point $\s \in \pa$ is
\begin{equation}
j_\Lambda(\x_0,\s;t) = \biggl( - D\frac{\partial}{\partial n} G_\Lambda(\x_0,\x;t)\biggr)_{\x=\s} = \frac{D}{\Lambda} G_\Lambda(\x_0,\s;t) ,
\end{equation}
where we used the Robin boundary condition (\ref{eq:G_Robin}).  This
is the joint probability density for the reaction time and reaction
point on the boundary (from the starting point $\x_0$).  The integral
over $\s$ yields the marginal probability density of reaction times,
\begin{equation}  \label{eq:rhot}
\rho_\Lambda(t;\x_0) = \int\limits_\pa d\s \, j_\Lambda(\x_0,\s;t) = \frac{D}{\Lambda} \int\limits_\pa d\s \, G_\Lambda(\x_0,\s;t) ,
\end{equation}
whereas the integral over time $t$ yields the marginal probability
density of reaction points:
\begin{equation}  \label{eq:omega}
\omega_\Lambda(\s;\x_0) = \int\limits_0^\infty dt \, j_\Lambda(\x_0,\s;t) = \frac{D}{\Lambda} \tilde{G}_\Lambda(\x_0,\s;p=0).
\end{equation}
The latter was called the {\it spread harmonic measure}
\cite{Grebenkov06a,Grebenkov06,Grebenkov06b,Grebenkov15}.  This is a
natural extension of the harmonic measure \cite{Garnett} that
characterizes the first arrival onto the perfectly reactive boundary
\cite{Grebenkov05a,Grebenkov05b}.

In general, the reactive surface does not need to be the whole
boundary of the confining domain $\Omega$.  The above analysis remains
applicable when only a part of the boundary is partially reactive,
whereas the remaining part is inert.  In this case, the Neumann
boundary condition is imposed on the inert part so that one faces the
mixed Robin-Neumann boundary condition.  In this setting, the
propagator $G_0(\x_0,\x;t)$ corresponds to the respective
Dirichlet-Neumann problem, with a perfectly reactive part.  In other
words, the presence of the inert part does not change the above
arguments, once the involved quantities are treated accordingly.

\section{An example: the half-space
}
\label{sec:half}

As a basic example, we consider the partially reflected Brownian
motion in an upper half-space, $\Omega = \{\x=(\y,z)\in \R^d ~:~ z >
0\}$, with partially reactive hyperplane $\pa = \R^{d-1}$.  For
convenience, we represent each point $\x$ as $(\y,z)$, where $\y \in
\R^{d-1}$ are lateral coordinates, and $z \geq 0$ is the transverse
coordinate.  In this example, the transverse and lateral motions are
independent, and the propagator $G_\Lambda(\x_0,\x;t)$ is the product
of the Gaussian propagator in lateral directions and the propagator on
the transverse positive semi-axis with the Robin boundary condition at
the zero endpoint.  The latter propagator is well known (see, e.g.,
the book \cite{Cole}), and we get
\begin{eqnarray}  \nonumber
G_\Lambda(\x_0,\x;t) &=& \frac{\exp\bigl(-\frac{|\y - \y_0|^2}{4Dt}\bigr)}{(4\pi Dt)^{(d-1)/2}}
\biggl\{\frac{\exp\bigl(- \frac{(z - z_0)^2}{4Dt}\bigr) + \exp\bigl(- \frac{(z + z_0)^2}{4Dt}\bigr)}{\sqrt{4\pi Dt}}  \\   \label{eq:G_space}
&& - \frac{1}{\Lambda} \exp\biggl(\frac{z + z_0}{\Lambda} + \frac{Dt}{\Lambda^2}\biggr) 
\erfc\biggl(\frac{z + z_0}{\sqrt{4Dt}} + \frac{\sqrt{Dt}}{\Lambda}\biggr) \biggr\} ,
\end{eqnarray}
where $\erfc(z)$ is the complementary error function.  This is a rare
example when the propagator can be obtained in a fully explicit form.
In the limit $\Lambda = 0$, Eq. (\ref{eq:G_space}) is reduced to the
Dirichlet propagator in the upper half-space:
\begin{equation} 
G_0(\x_0,\x;t) = \underbrace{\frac{\exp\bigl(-\frac{|\y - \y_0|^2}{4Dt}\bigr)}{(4\pi Dt)^{(d-1)/2}}}_{\textrm{lateral}} \,
\underbrace{\frac{\exp\bigl(-\frac{(z-z_0)^2}{4Dt}\bigr) - \exp\bigl(- \frac{(z+z_0)^2}{4Dt}\bigr)}{\sqrt{4\pi Dt}}}_{\textrm{transverse}} \, ,
\end{equation}
from which
\begin{equation}
j_0(\x_0,\s;t) = \frac{\exp\bigl(-\frac{|\s - \y_0|^2}{4Dt}\bigr)}{(4\pi Dt)^{(d-1)/2}} \, \frac{z_0 \exp\bigl(-\frac{z_0^2}{4Dt}\bigr)}{\sqrt{4\pi D t^3}} \,.
\end{equation}

\begin{figure}
\begin{center}
\includegraphics[width=67mm]{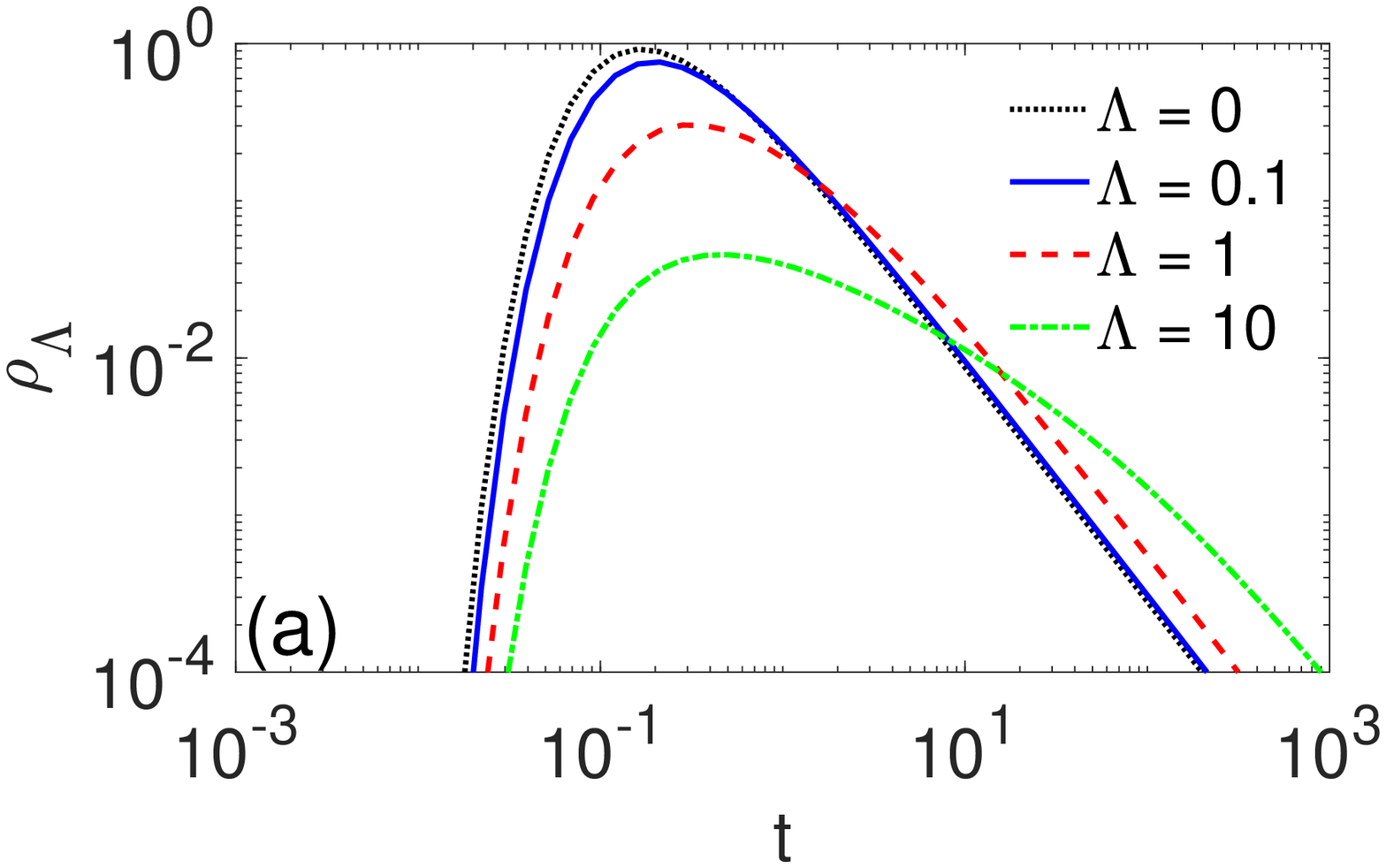}
\includegraphics[width=67mm]{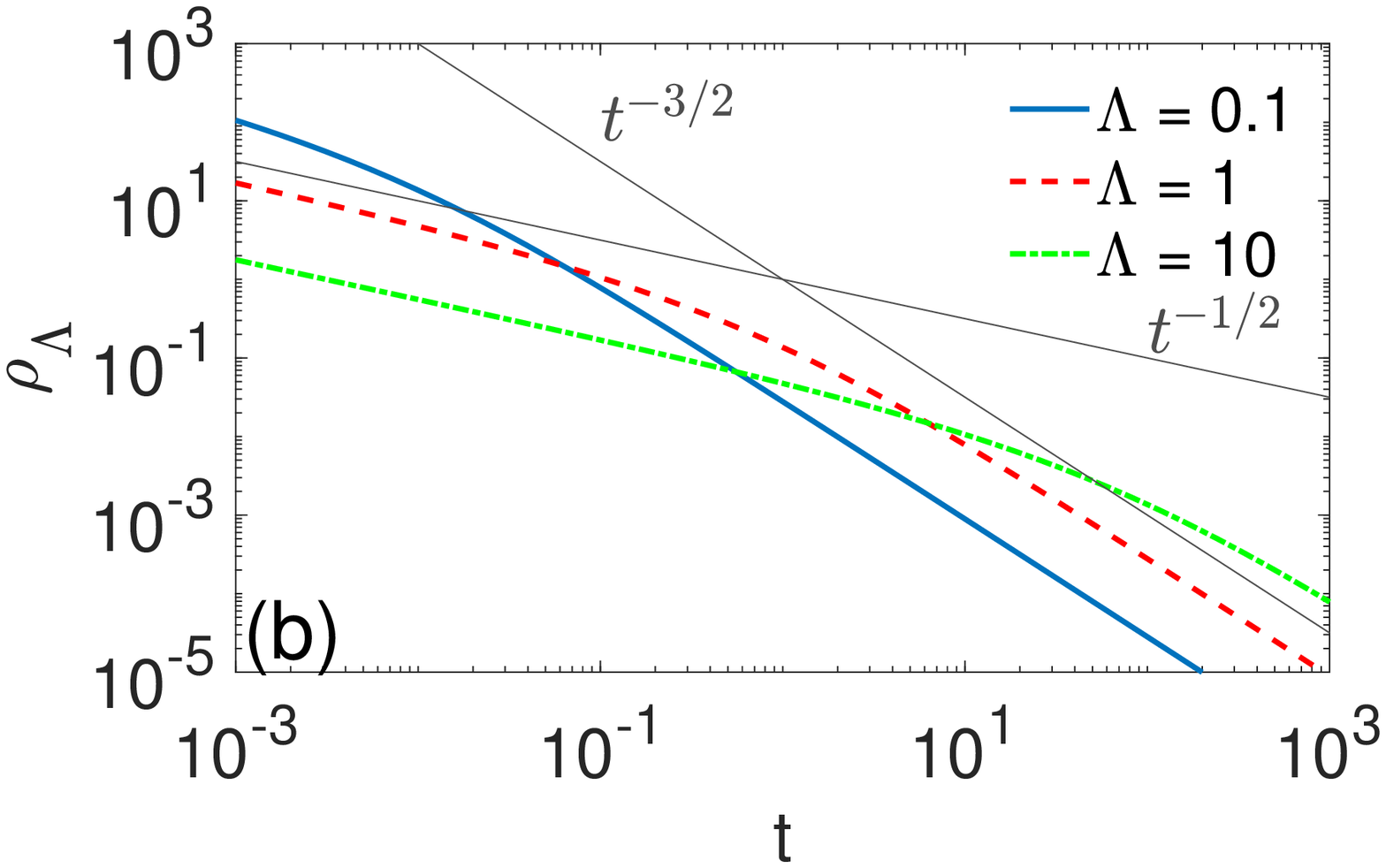}
\end{center}
\caption{
The probability density $\rho_\Lambda(t;z_0)$ of the reaction time for
a molecule started at height $z_0$ above a flat surface, with $z_0 =
1$ {\bf (a)} and $z_0 = 0$ {\bf (b)}, and four values of the reaction
length $\Lambda$ (arbitrary units are used, with $D = 1$).  In the
second plot, the density $\rho_0(t;z_0=0) = \delta(t)$ for $\Lambda =
0$ is not shown.  Gray lines indicate power law asymptotics $t^{-1/2}$
and $t^{-3/2}$.}
\label{fig:rho}
\end{figure}

The above formulas help to compute other quantities of interest.  For
instance, the probability density of the reaction time follows from
Eqs. (\ref{eq:rhot}) 
\begin{equation}
\rho_\Lambda(t;z_0) =  \frac{D\exp\bigl(-\frac{z_0^2}{4Dt}\bigr)}{\Lambda}\biggl\{\frac{1}{\sqrt{\pi Dt}} 
 - \frac{1}{\Lambda} \, \erfcx\biggl(\frac{z_0}{\sqrt{4Dt}} + \frac{\sqrt{Dt}}{\Lambda}\biggr) \biggr] \biggr\} 
\end{equation}
(note that the density depends only on the height $z_0$ of the
starting point $\x_0 = (\y_0,z_0)$ above the boundary).  In the limit
$\Lambda = 0$, one retrieves the classical expression $\rho_0(t;z_0) =
\frac{z_0}{\sqrt{4\pi D t^3}} e^{-z_0^2/(4Dt)}$.  In the long-time limit,
one gets $\rho_\Lambda(t;z_0) \simeq \frac{z_0+\Lambda}{\sqrt{4\pi
Dt^3}}$, as though the height of the starting point is increased by
the reaction length $\Lambda$.  In the short-time limit, one gets
either a very fast vanishing as $t\to 0$ for $z_0 > 0$, or a power law
divergence $\rho_\Lambda(t;z_0) \simeq \frac{D/\Lambda}{\sqrt{\pi D
t}}$ for $z_0 = 0$.  This behavior is illustrated in
Fig. \ref{fig:rho}.

\begin{figure}
\begin{center}
\includegraphics[width=67mm]{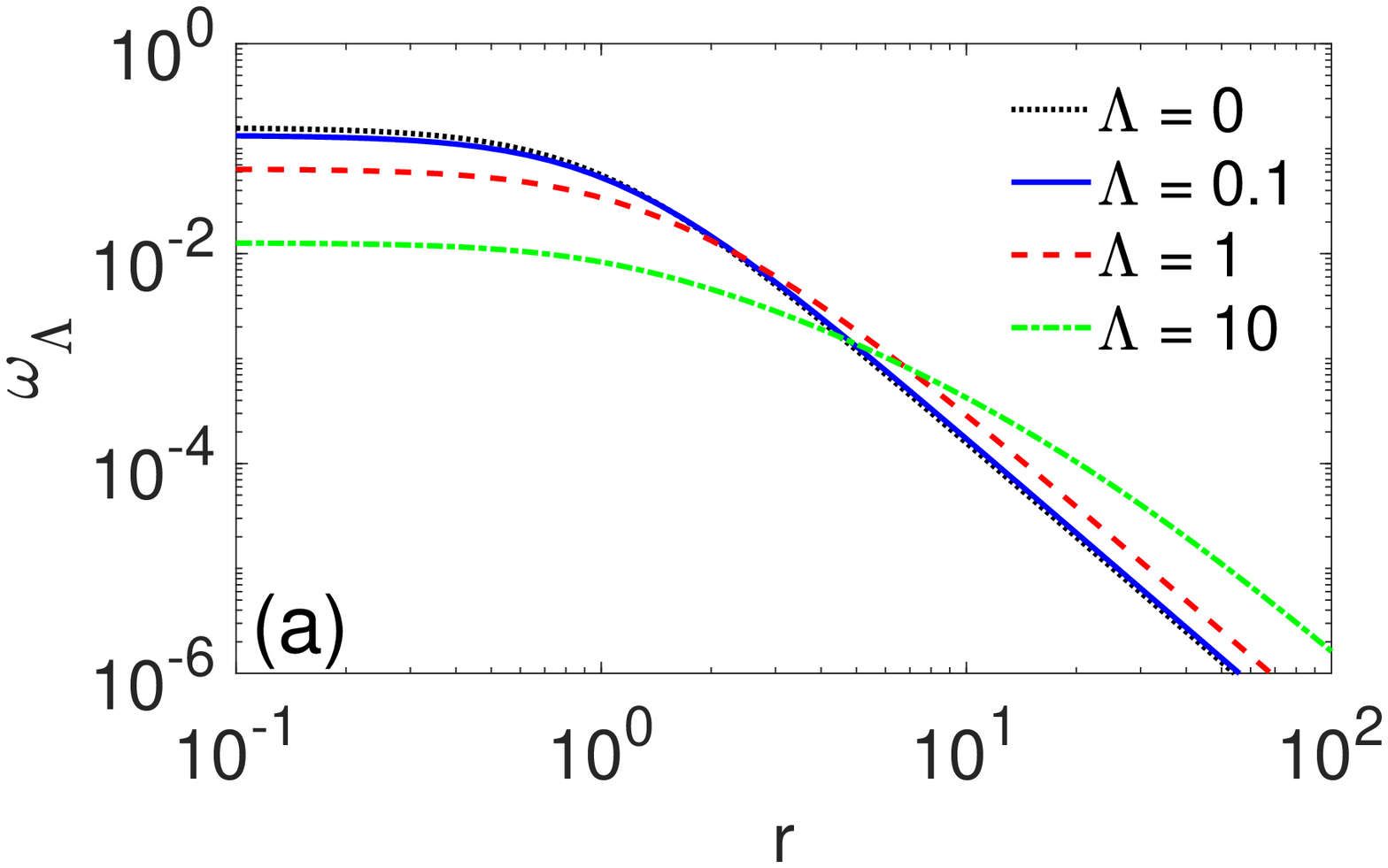}
\includegraphics[width=67mm]{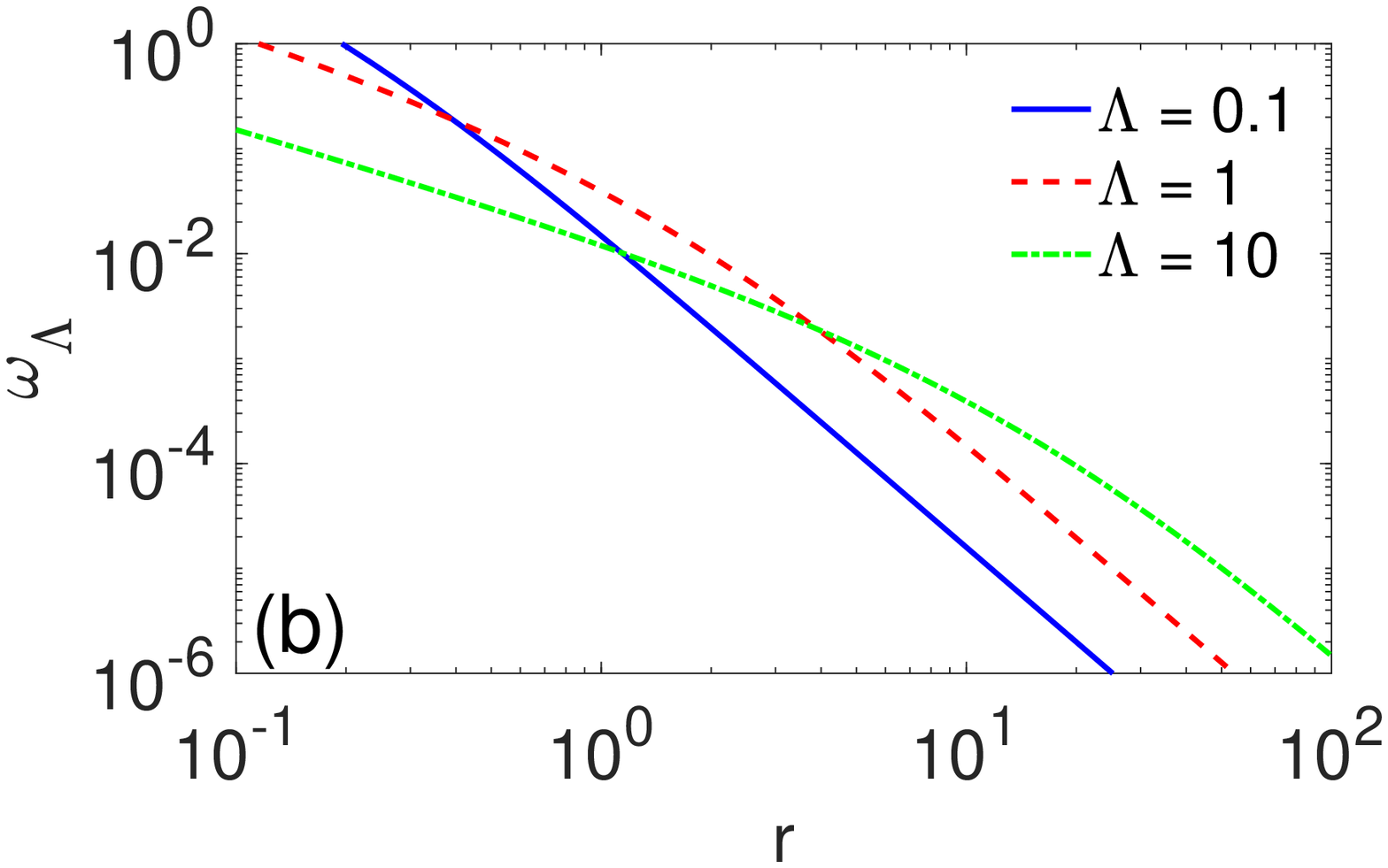}
\end{center}
\caption{
The spread harmonic measure density $\omega_\Lambda(\s;(\y_0,z_0))$ of
the reaction point $\s$ as a function of the radial distance $r = |\s
- \y_0|$ for a molecule started at height $z_0$ above the plane in the
three-dimensional space ($d = 3$), with $z_0 = 1$ {\bf (a)} and $z_0 =
0$ {\bf (b)}, and four values of the reaction length $\Lambda$
(arbitrary units are used, with $D = 1$).  In the second plot, the
harmonic measure density $\omega_0(\s;(\y_0,0)) = \delta(\s-\y_0)$ for
$\Lambda = 0$ is not shown. }
\label{fig:omega}
\end{figure}

The spread harmonic measure density is obtained from
Eq. (\ref{eq:omega}): 
\begin{equation}  \label{eq:omega_half}
\omega_\Lambda(\s;\x_0) = \int\limits_{\R^{d-1}} \frac{d\q}{(2\pi)^{d-1}} \, \frac{e^{-i(\q (\s-\y_0))}}{1 + \Lambda |\q|} =
\int\limits_0^\infty dz \frac{e^{-z/\Lambda}}{\Lambda} \, \omega_0(\s;(\y,z_0+z)) ,
\end{equation}
where
\begin{equation}  \label{eq:HM}
\omega_0(\s;(\y_0,z_0)) = \frac{\Gamma(d/2)}{\pi^{d/2}} \, \frac{z_0}{[z_0^2 + |\s - \y_0|^2]^{d/2}} 
\end{equation}
is the harmonic measure density (see Ref. \cite{Grebenkov15} for the
derivation of the last equality in Eq. (\ref{eq:omega_half})).  In
other words, the effect of partial reflections on the boundary can be
seen as an effective increase the height $z_0$ of the starting point
above the boundary.  This increase is randomly distributed with an
exponential law determined by the reaction length $\Lambda$.  As
expected, the density $\omega_\Lambda$ is radially symmetric with
respect to the line which is perpendicular to the reactive surface and
passes through the starting point $\x_0$.  In other words, it depends
only on the height $z_0$ and the radial distance $r = |\s - \y_0|$.
The behavior of the spread harmonic measure is illustrated in
Fig. \ref{fig:omega}.  One can easily check that the {\it mean}
exploration distance, defined as the standard deviation of explored
distance from the first arrival point to the reaction point, is
infinite (see also Ref. \cite{Sapoval05}).  This is related to the
fact that each jump can go unlimitedly far away from the reactive
boundary, and such rare but anomalously long trajectories dominate in
the second moment, due to the power-law asymptotic decay of the
harmonic measure density in Eq. (\ref{eq:HM}).  In this way, the
successive arrival points of partially reflected Brownian motion onto
the reactive surface can be seen as L\'evy flights on that surface
\cite{Mandelbrot,Viswanathan96,Levitz05,Levitz06,Levitz13}.

\begin{figure}
\begin{center}
\includegraphics[width=80mm]{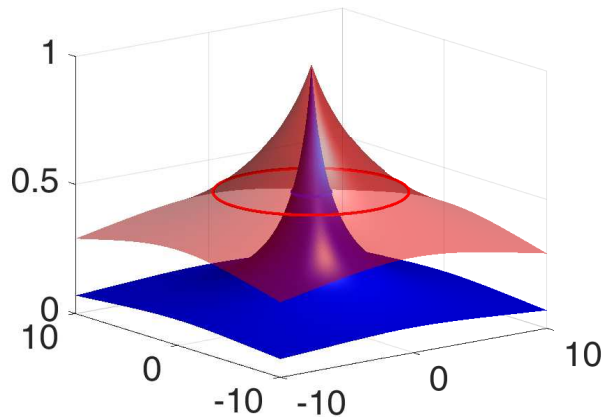}
\end{center}
\caption{
The probability $P_\Lambda(R)$ of the reaction event at a boundary
point $\s$ at distance greater than $R$ to the first arrival point
(here, the origin) on the plane in the three-dimensional space
($d=3$), for two values of the reaction length: $\Lambda = 1$ (blue
narrower surface) and $\Lambda = 5$ (red broader surface).  Two
circles show the median reaction radius $R_m \approx 1.17 \, \Lambda$
at which this probability is equal to $0.5$.  Arbitrary length units
are used.}
\label{fig:P}
\end{figure}

Integrating the spread harmonic measure over $\s$ with $|\s - \y_0| >
R$, one gets the probability of the reaction event at boundary points
whose distance to the first arrival point is greater than $R$:
\begin{equation}
P_\Lambda(R) = \int\limits_{|\s - \y_0|>R} d\s \, \omega_\Lambda(\s;(\y_0,0)) = \frac{2\Gamma(d/2)}{\sqrt{\pi}\, \Gamma((d-1)/2)} 
\int\limits_0^\infty dx \, \frac{e^{-xR/\Lambda}}{(1 + x^2)^{d/2}} 
\end{equation}
(the last equality was derived in Ref. \cite{Grebenkov15}).  This
probability characterizes the exploration of a flat partially reactive
surface after the first arrival and till the reaction.  This is a
function of $R/\Lambda$, which monotonously decreases from $1$ to $0$.
The condition $P_\Lambda(R_m) = 0.5$ defines the median radius $R_m$
of the spherical domain around the first arrival point, at which half
of molecules react.  In three dimensions, one gets $R_m \approx 1.17
\, \Lambda$, providing a geometric interpretation of the reaction
length $\Lambda$.  The behavior of the probability $P_\Lambda(R)$ is
illustrated in Fig. \ref{fig:P}.

\section{Conclusion: Is the finite reactivity important?}

As discussed in Sec. \ref{sec:intro}, there are various microscopic
mechanisms that naturally lead to a finite reactivity of the target
surface.  However, the mathematical description of imperfect reactions
involves more sophisticated Robin boundary condition and partially
reflected Brownian motion.  While most of the results known for
perfect reactions can be generalized to imperfect ones, mathematical
derivations are usually much more involved.  It is probably for this
reason that the majority of theoretical studies remains focused on
perfect reactions.  It is therefore natural to ask whether the finite
reactivity is indeed important?

The reactivity of the target introduces the reaction length $\Lambda =
D/\kappa$ which controls the balance between the transport step toward
the target and the reaction step on the target.  The latter step
involves the intricate exploration of the partially reactive surface
until the reaction occurs.  This process mixes bulk diffusion from
surface to surface points and eventual reaction trials.  Since the
process is strongly coupled to the geometry of the environment, the
dependence of diffusion characteristics (such as the total flux) on
$\Lambda$ can reveal geometric structure of the target surface
\cite{Filoche08}.

When the reaction length is the smallest length scale of the problem,
the effects of finite reactivity can be neglected.  In practice,
however, there is a conceptual difference between setting $\Lambda =
0$ (the idealized case of perfect reactions with Dirichlet boundary
condition, $\kappa = \infty$), and $\Lambda > 0$, even if $\Lambda$
remains small.  For instance, in the analysis of the short-time
asymptotic behavior of a diffusion problem, the diffusion length
$\sqrt{Dt}$ can become smaller than the reaction length $\Lambda > 0$
as $t\to 0$.  As a consequence, the short-time asymptotic behavior for
perfect and imperfect reactions is different
\cite{Grebenkov10a,Grebenkov10b}.  For example, the probability
density of the reaction time on a spherical target of radius $R$ reads
\cite{Grebenkov10a}
\begin{equation}
\rho_\Lambda(t;\x_0) \simeq \frac{R}{|\x_0|} \, \frac{\exp\bigl(-\frac{(|\x_0|-R)^2}{4Dt}\bigr)}{t \, \sqrt{\pi}} \times
\begin{cases} \displaystyle \frac{|\x_0|-R}{\sqrt{4Dt}} \qquad (\Lambda = 0) , \cr 
\displaystyle   \frac{\sqrt{Dt}}{\Lambda} \hskip 13mm  (\Lambda > 0)  \end{cases}
\end{equation}
(with the starting point $\x_0$ outside the target: $|\x_0| > R$).
One can see that the power law prefactor is different in two cases.
Moreover, when $\Lambda$ is small, one can also expect the transition
from the short-time regime at $\sqrt{Dt} \ll \Lambda \ll |\x_0|-R$
(given by the second expression) to an intermediate regime at $\Lambda
\ll \sqrt{Dt} \ll |\x_0|-R$ (given by the first expression).

Another important example is the so-called narrow escape problem from
an Euclidean domain $\Omega$ through a small hole $\Gamma$ on the
boundary $\pa$ (see the review \cite{Holcman14} and references
therein).  In this setting, the escape through a hole $\Gamma$ can be
understood as reaction on the target $\Gamma$ upon the first arrival.
When the (nondimensional) size $\ve$ of the escape region goes to
zero, the mean exit time diverges as $\log(1/\ve)$ in two dimensions
and as $\ve^{-1}$ in three dimensions \cite{Singer06,Holcman14}.
However, an escape through a small opening requires overpassing either
an entropic or an energetic barrier, or both.  Once the escape region
is considered as partially reactive ($\Lambda > 0$), the asymptotic
behavior of the mean exit time changes drastically
\cite{Grebenkov17a}.  In the narrow escape limit $\ve \to 0$, the
dominant contribution to the mean exit time comes from the finite
reactivity and scales as $\ve^{-1}/\kappa$ in two dimensions and
$\ve^{-2}/\kappa$ in three dimensions.  This contribution does not
exist for idealized perfect escape ($\kappa = \infty$) but becomes
dominant for a finite reactivity $\kappa$.  Once again, in the narrow
escape limit, the size of the hole becomes the smallest length scale,
and thus the finite reactivity cannot be ignored.  The impact of
$\kappa$ on the whole distribution of the reaction time was recently
investigated \cite{Grebenkov18}.

These two examples illustrate how an idealized assumption of perfect
reactions can be misleading.  Even though the analysis of imperfect
reactions is mathematically more involved, the related technical
difficulties are generally not insurmountable.  As a matter of fact,
the finite reactivity often ``regularizes'' solutions and helps to
resolve some apparent ``paradoxical'' properties of Brownian motion.
For instance, the Smoluchowski flux (\ref{eq:J_Smol}) toward a
perfectly reactive spherical target diverges as $t\to 0$, meaning that
too many molecules in a close vicinity of the target react at first
time instances.  This unrealistic divergence does not appear in the
Collins-Kimball flux (\ref{eq:J_Collins}) toward a partially reactive
target.  Moreover, partially reflected Brownian motion can be started
right at the partially reactive boundary whereas ordinary Brownian
motion started on an absorbing boundary would immediately react.  As a
consequence, the Robin boundary condition is necessary for a natural
implementation of reversible association of a diffusing molecule with
the target.  In fact, when the formed metastable complex $AB$ in
(\ref{eq:reaction2}) dissociates (with some rate $k_{\rm off}$), the
just dissociated molecule is released {\it on} the target surface.  If
the latter was perfectly reactive, the molecule would {\it
immediately} react again, making such a dissociation process
impossible \cite{Prustel13,Grebenkov17c}.  A standard trick to
overcome this problem consists in ejecting the just dissociated
molecule to a finite distance $a$ from the target, from which its
diffusion is released
\cite{Filoche99,Grebenkov03,Benichou10,Benichou11,Rupprecht12a,Rupprecht12b}.
While the introduction of such an ejection distance can be
rationalized from the physical point of view (e.g., a finite size of
molecules, the presence of a specific surface layer, a finite range
binding interaction, etc.), this is precisely an intermediate step of
the construction of partially reflected Brownian motion as described
in Sec. \ref{sec:proba}.  In other words, even when one tries to avoid
using partial reactivity and the associated PRBM, one often uses it
implicitly via a ``regularization'' by a finite ejection distance $a$.

The mathematical theory of partially reflected diffusions is rather
well developed \cite{Ma90,Papanicolaou90,Milshtein95,Bass08,Schuss}.
In particular, the fundamental interconnection between the stochastic
process, the diffusion equation with Robin boundary condition, and the
spectral properties of the underlying diffusion operator and the
Dirichlet-to-Neumann operator provides the solid mathematical
foundation for investigating imperfect diffusion-controlled reactions.
In turn, the intricate exploration of the partially reactive surface
via diffusion-mediated jumps in complicated structures such as
multiscale porous media or domains with irregular or fractal
boundaries, remains poorly understood.  In this light, efficient
numerical techniques such as fast Monte Carlo methods
\cite{Kansal02,Grebenkov05a,Grebenkov05b,Grebenkov14c} or
semi-analytical solutions \cite{Galanti16,Grebenkov18b} become
particularly important.

\end{document}